\documentclass[aps,pra,amsmath,amssymb,11pt,final,tightenlines,twoside,onecolumn,nofloats,nofootinbib,superscriptaddress,
showkeys,showkeywords]{revtex4-2}

\usepackage[T2A]{fontenc}
\usepackage[utf8x]{inputenc}
\usepackage[russian,english]{babel}
\usepackage{graphicx}
\usepackage{longtable}

\input{maik_edt_en.rty}

\def\aj{Astron. J}
\def\apj{Astrophys. J}
\def\apjl{Astrophys. J. Let.}

\def\aap{Astron. Astrophys.}

\def\mnras{MNRAS}

\def\pasj{PASJ}
\def\pasp{PASP}

\def\sovast{Soviet Astronomy}
\def\na{New Astronomy}
\def\actaa{Acta Astronomica}

\begin{document}

\noindent {\it ASTRONOMY REPORTS, 2026, Vol. , № }
\bigskip\bigskip  \hrule\smallskip\hrule
\vspace{35mm}


\title{WOLF-RAYET STARS AND BLACK HOLES: MASS DISTRIBUTIONS AND EVOLUTIONARY CONNECTION}

\author{\bf\copyright~2026 г. 
\firstname{I.~A.}~\surname{Shaposhnikov}
}
\email{iv.shaposhnikov@gmail.com}
\author{
\firstname{K.~A.}~\surname{Postnov}
}
\author{
\firstname{A.~M.}~\surname{Cherepashchuk}
}
\affiliation{Lomonosov Moscow State University, Sternberg Astronomy Institute, Moscow, 119234, Russia}

\begin{abstract}
\vspace{3mm}
\received{}
\revised{}
\accepted{} 
\vspace{3mm}

The form and relationship between the mass distributions of Wolf-Rayet stars and black holes are analyzed using a direct method for estimating the distribution function from a sample of mass values with individual distributions obtained from current observations. It is shown that, taking into account the mass loss of WR stars by the end of this evolutionary stage, the mass distribution of black holes is close to that of the CO-cores of WR stars before gravitational collapse, $M_\mathrm{BH}\simeq M_\mathrm{CO}$.

\end{abstract}
\maketitle

\section{Introduction}

The final stage in the evolution of a massive star ($M_\mathrm{ZAMS} \gtrsim 8 \div 10~M_\odot$) is the gravitational collapse of its core resulting in the formation of a compact object — a neutron star (NS) or a black hole (BH). Shortly before the core collapse, a massive star, having by then lost most of its outer hydrogen envelope, is at the Wolf-Rayet star (WR) stage. In the direct gravitational collapse of the carbon-oxygen (CO) core of a WR star with a mass exceeding the Oppenheimer-Volkoff limit ($M_\mathrm{CO} > M_\mathrm{OV} \approx 3~M_\odot$), the black hole formation is the most likely outcome. In close binary systems (CBS) with black holes observed as high-mass or low-mass X-ray binaries (HMXB/LMXB), black hole formation should also proceed through a CBS stage with a WR star \cite{1973NPhS..242...71V,1973NInfo..27...70T,2014LRR....17....3P}.

To date, dozens of mass estimates have been obtained for WR and BH stars predominantly residing in spectroscopic and X-ray binaries. This enables a direct comparison of the stellar mass distributions at two successive stages of their evolution — before and after the core collapse. A similar comparison, performed for the WR and BH mass distributions, demonstrated a close relation between black hole masses and the mass of CO-cores in pre-collapse stellar stars \cite{WRBH,WRCO}. This paper aims to provide a more detailed analysis of the form and relationship of these distributions using a direct method for estimating the distribution function from a sample of mass values with observed individual distributions, without making a priori assumptions about the parametric form of the distribution.

\section{Observational data} 

\subsection{WR masses from WR+OB binaries}

The analysis of the WR and BH mass distributions in this paper is based on data similar to those used in our previous paper \cite{WRBH}.
These data include mass estimates for 31 Wolf-Rayet stars from WR+OB spectroscopic binaries in our Galaxy and the Magellanic Clouds. Only binaries with reliably measured radial velocity curves for both stars were used in the analysis. Compared to \cite{WRBH}, here we have added data for the recently discovered LS III +44 21 system \cite{arXiv2605.24257} to the WR sample. We have also refined the masses of WR stars for a number of systems for which we carried out spectroscopic observations (see below). In these refinements, the same data were used as in the previous studies, but the technique from \cite{arXiv2605.24257} was employed: based on the entire set of stellar radial velocity measurements in the lines of various ions, an optimization was performed using the half-amplitudes of the radial velocities $K_\mathrm{WR}$, $K_\mathrm{OB}$ and a set of $\gamma$-velocities independent for each line (i.e., systematic velocities for each radial velocity curve). This technique allows us to correctly take into account the observed spread of $\gamma$-velocities and reduces the error in determining $K_\mathrm{WR}$ and  $K_\mathrm{OB}$ from which the stellar masses are derived with an accuracy of up to a factor of $\sin^3i$ ($i$ is the binary inclination angle to the line of sight). Refining masses of these binaries is crucial for the present study because we estimate the parameters of the empirical dependence of the mass-loss rate of WR stars on their masses for these systems based on the observed evolution of their orbital periods (see Section \ref{sec:dot_MWR}). The refined semi-amplitudes of the radial velocities and masses of these WR stars (to an accuracy of $\sin^3i$) are presented in Table \ref{tab:WR_enchanced}.

\begin{table}[h]
    \centering
    \caption{Refined parameters of spectroscopic orbits for the binaries WR 127 = Hen~3-1772, WR 151 = CX~Cep, WR 141 = V2183~Cyg, WR 139 = V444~Cyg and WR 155 = CQ Cep. When calculating the component masses in the WR 155 system, the mass ratio $q = M_\mathrm{{WR}}/M_\mathrm{{OB}} = 0.58\pm0.03$ from paper \cite{Kartashova1985} was adopted. The last column provides references to radial velocity curves we use.}
    \label{tab:WR_enchanced}
    \begin{tabular}{|c|c|c|c|c|c|}
         \hline
         System & $K_\mathrm{WR}$, km/s & $K_\mathrm{OB}$, km/s &  $M_\mathrm{{WR}}\sin^3i,~M_\odot$ & $M_\mathrm{{OB}}\sin^3i,~M_\odot$ & Ref. \\
         \hline
         WR 127 & 184.5$\pm$1.3 &  98$\pm$4 & 7.8$\pm$1.4 & 14.6$\pm$1.5 & \cite{2024AnA...683L..17S} \\
         WR 151 & 326.7$\pm$0.7 & 159.6$\pm$1.8 & 8.3$\pm$0.5 & 17.1$\pm$0.5 & \cite{2023MNRAS.523.1524S} \\
         WR 141 & 129.3$\pm$2.6 & 138$\pm$6 & 22.2$\pm$1.9 & 20.8$\pm$1.5 & \cite{2024ARep...68.1145S} \\
         WR 139 & 307.9$\pm$1.0 & 121.5$\pm$1.9 & 9.8$\pm$1.1 & 24.8$\pm$1.4 & \cite{2023MNRAS.526.4529S} \\
         WR 155 & 297.4$\pm$1.1 & -- & 6.5$\pm$0.4 & 11.2$\pm$0.4 & \cite{2023MNRAS.523.1524S} \\
         \hline
    \end{tabular}
\end{table}

Note that in addition to WR+OB  close binary systems ($P_{orb} \lesssim 30$ days) we also include several WR star masses in wide binary systems  ($P_{orb} \approx 71\div1500$ days) inferred from long-term interferometric observations \cite{2021ApJ...908L...3R,2024ApJ...977..185H,2007MNRAS.377..415N,2024ApJ...977...78R,2021MNRAS.504.5221T}.

\begin{figure}[h]
    \centering
    \includegraphics[width=0.9\linewidth]{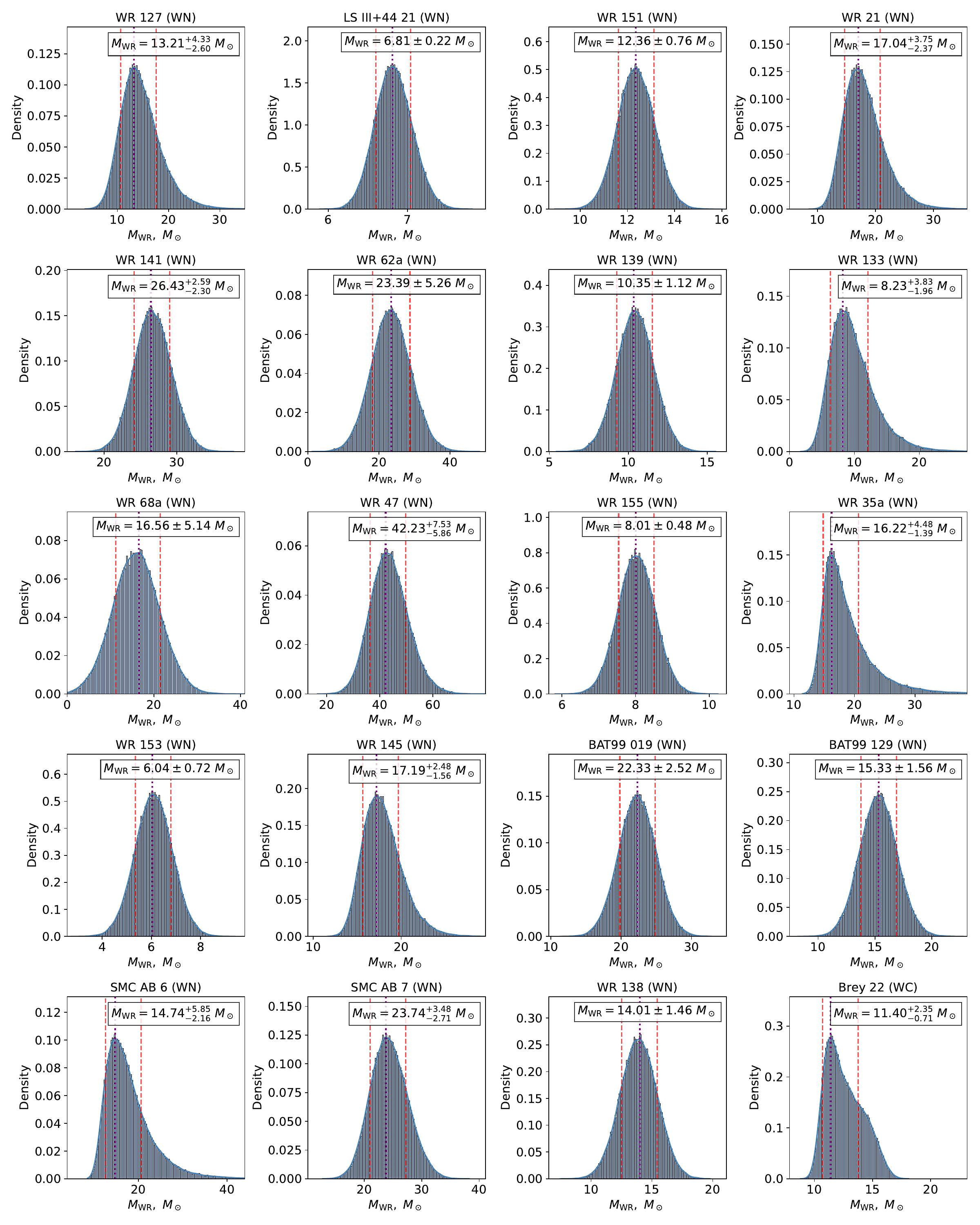}
    \caption{Individual mass distributions of WR stars for WR+OB binary systems from Table \ref{tab:MWR}.}
    \label{fig:WR_individuals_1}
\end{figure}
\begin{figure}[h]
    \centering
    \includegraphics[width=0.9\linewidth]{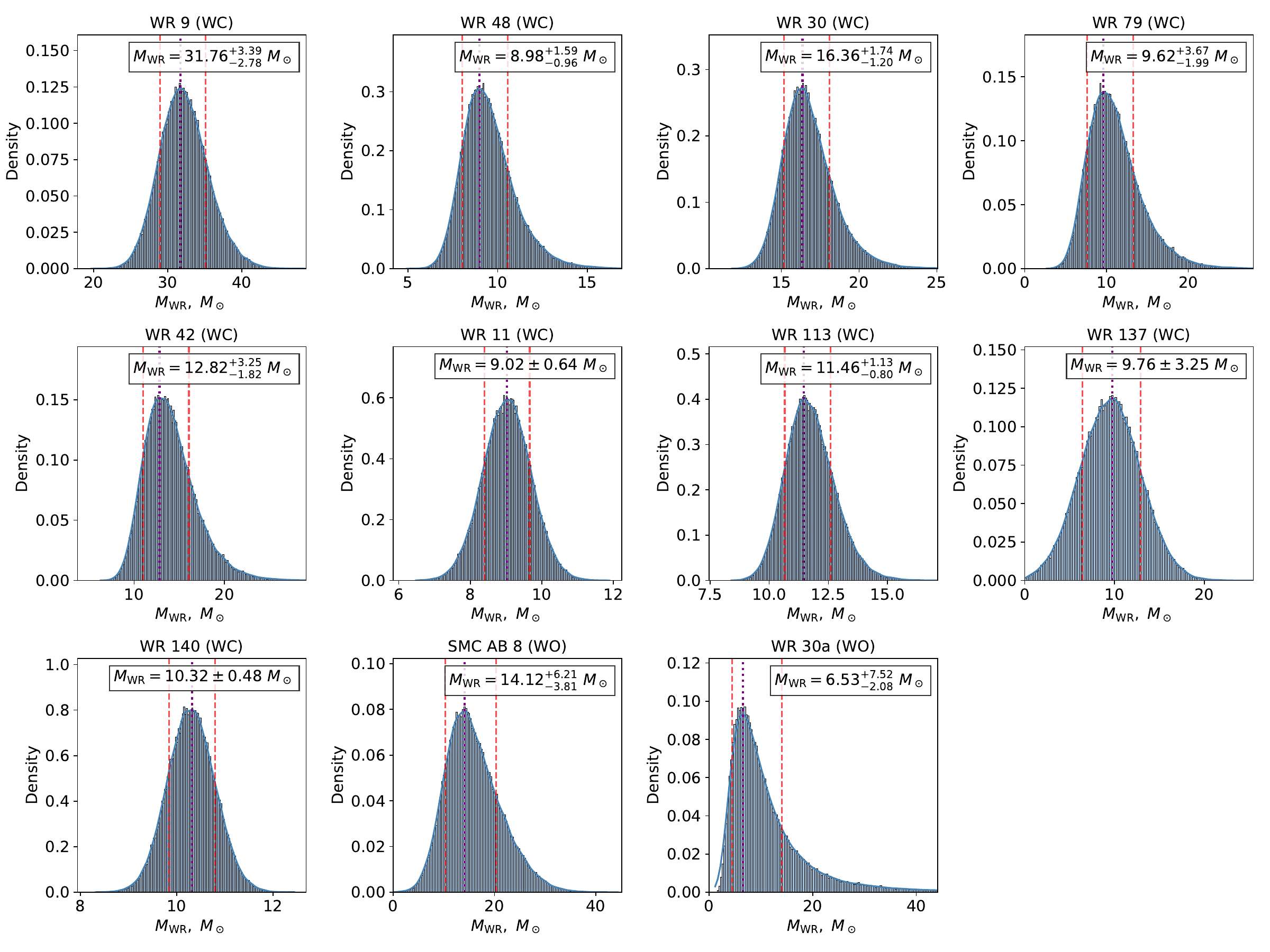}
    \caption{Individual mass distributions of WR stars for WR+OB binary systems from Table \ref{tab:MWR} (continued).}
    \label{fig:WR_individuals_2}
\end{figure}

For all systems selected for the analysis, we calculated individual $M_\mathrm{{WR}}$ distribution by generating $10^5$ values of $M_\mathrm{{WR}}$ based on an artificial sample of $M_\mathrm{{WR}}\sin^3i$ and $i$ generated from their observed distributions. After that the resulting $M_\mathrm{{WR}}$ distribution was smoothed with a Gaussian kernel. We assumed normal distribution for $M_\mathrm{{WR}}\sin^3i$ and normal or uniform distributions for $i$, depending on the observational constraints. The individual WR mass distributions $M_\mathrm{{WR}}$ obtained in this way are presented in Figs. \ref{fig:WR_individuals_1} and \ref{fig:WR_individuals_2}.

\break

Information on WR stars used in this paper is summarized in Table \ref{tab:MWR}. For each system, we provide $M_\mathrm{{WR}}\sin^3i$ and $i$ used in calculating the mass estimate $M_\mathrm{{WR}}$ and its error from the generated distribution. The symmetric uncertainties in $M_\mathrm{{WR}}\sin^3i$ and $i$ marked with $\pm$ sign mean the normal error distribution. For quantities whose distribution is assumed to be uniform, the smallest and largest values are given separated by the $\div$ sign.

\begin{longtable}{|c|c|c|c|c|c|}
    \caption{Observational estimates of masses of Wolf-Rayet stars.}
    \label{tab:MWR}
    \\ \hline System & WR type & $M_\mathrm{{WR}}\sin^3i,~M_\odot$ & $i,~^\circ$ & $M_\mathrm{{WR}},~M_\odot$ & Ref. \\  \hline \endfirsthead
    
    \caption{Observational estimates of masses of Wolf-Rayet stars (continued).}
    \\ \hline System & WR type & $M_\mathrm{{WR}}\sin^3i,~M_\odot$ & $i,~^\circ$ & $M_\mathrm{{WR}},~M_\odot$ & Ref. \\  \hline \endhead
    
    \hline  \endfoot
    \hline \endlastfoot
        WR 127 & WN & $7.8\pm1.4$ & $55\pm5$ & $13_{-3}^{+4}$ & \cite{2024AnA...683L..17S,1996AJ....112.2227L} \\
        LS III +44 21 & WN & $6.69\pm0.19$ & $79\div90$ & $6.81\pm0.22$ & \cite{arXiv2605.24257} \\
        WR 151 & WN & $8.3\pm0.5$ & $61.1\pm0.7$ & $12.4\pm0.8$ & \cite{2023MNRAS.523.1524S,2009PASP..121..708H} \\
        WR 21 & WN & $7.8\pm0.6$ & $49.6\pm3.7$ & $17.0_{-2.4}^{+4}$ & \cite{2012MNRAS.424.1601F,1996AJ....112.2227L} \\
        WR 141 & WN & $22.2\pm1.9$ & $71.7\pm1.4$ & $26.6\pm2.4$ & \cite{2024ARep...68.1145S,1996AJ....112.2227L} \\
        WR 62a & WN & $22\pm5$ & $70\div90$ & $23\pm5$ & \cite{2013AnA...552A..22C} \\
        WR 139 & WN & $9.8\pm1.1$ & $78.3\pm0.4$ & $10.4\pm1.2$ & \cite{2023MNRAS.526.4529S,2011AN....332..616E}  \\
        WR 133 & WN & $0.27\pm0.05$ & $17.9\pm1.7$ & $8.2_{-2.0}^{+4}$ & \cite{2021ApJ...908L...3R} \\
        WR 68a & WN & $15\pm5$ & $80\pm5$ & $16\pm5$ & \cite{2015AnA...581A..49C} \\
        WR 47  & WN & $33.7\pm4.7$ & $67\pm3$ & $43\pm6$ & \cite{2012MNRAS.424.1601F} \\
        WR 155 & WN & $6.5\pm0.4$ & $68.8\pm0.6$ & $8.0\pm0.5$ & \cite{1997AN....318..267D,2023MNRAS.523.1524S} \\
        WR 35a & WN & $14.9\pm1.0$ & $71\pm10$ & $16.2_{-1.4}^{+5}$ & \cite{2014AnA...562A..13G} \\
        WR 153 & WN & $5.7\pm0.7$ & $78.0\pm1.0$ & $6.0\pm0.7$ & \cite{2002ApJ...577..409D,1996AJ....112.2227L} \\
        WR 145 & WN & $12.5\pm0.7$ & $63\pm4$ & $17.2_{-1.6}^{+2.5}$ &  \cite{2009MNRAS.399.1977M} \\
        BAT99 019 & WN & $22.1\pm2.6$ & $82\div90$ & $22\pm3$ & \cite{2019AnA...627A.151S} \\
        BAT99 129 & WN & $14.3\pm1.5$ & $78.0\pm2.0$ & $15.3\pm1.6$ & \cite{bat129-ph,2019AnA...627A.151S} \\
        SMC AB 6 & WN & $9.2\pm0.5$ & $57\pm8$ & $14.7_{-2.2}^{+6}$ & \cite{2018AnA...616A.103S} \\
        SMC AB 7 & WN & $18.0\pm2.0$ & $60\div70$ & $24\pm3$ & \cite{2002MNRAS.333..347N} \\
        WR 138 & WN & $13.7\pm1.5$ & $84.21\pm0.06$ & $14.0\pm1.5$ & \cite{2024ApJ...977..185H} \\

        Brey 22 & WC & $10.2\pm0.5$ & $60\div80$ & $11.4_{-0.7}^{+2.4}$ & \cite{1990ApJ...348..232M} \\
        WR 9 & WC & $18.8\pm1.4$ & $56.8\pm2.0$ & $32\pm3$ & \cite{1990ApJ...348..232M,1996AJ....112.2227L} \\
        WR 48 & WC & $7.7\pm0.7$ & $70\pm6$ & $9.0_{-1.0}^{+1.6}$ & \cite{2002MNRAS.335.1069H} \\
        WR 30 & WC & $15.4\pm1.0$ & $78\pm6$ & $16.4_{-1.2}^{+1.7}$ & \cite{2012MNRAS.424.1601F,1996AJ....112.2227L} \\
        WR 79 & WC & $1.8\pm0.3$ & $33.6\pm2.9$ & $9.6_{-2.0}^{+4}$ & \cite{1990ApJ...348..232M,1996AJ....112.2227L} \\
        WR 42 & WC & $3.7\pm0.3$ & $40.3\pm2.9$ & $12.8_{-1.8}^{+3}$ & \cite{1990ApJ...348..232M,1996AJ....112.2227L} \\
        WR 11 & WC & $6.8\pm0.5$ & $65.5\pm0.4$ & $9.0\pm0.6$ & \cite{2007MNRAS.377..415N} \\
        WR 113 & WC & $9.1\pm0.5$ & $67\pm3$ & $11.7\pm0.9$ & \cite{2018MNRAS.474.2987H,1996AJ....112.2227L} \\
        WR 137 & WC & $9.3\pm3.3$ & $82.86\pm0.06$ & $10\pm3$ & \cite{2024ApJ...977...78R} \\
        WR 140 & WC & $6.75\pm0.29$ & $60.3\pm0.7$ & $10.3\pm0.5$ & \cite{2021MNRAS.504.5221T} \\

        SMC AB 8 & WO & $5.0\pm1.3$ & $37\div50$ & $14_{-4}^{+6}$ & \cite{2016AnA...591A..22S} \\
        WR 30a & WO & $0.69\pm0.12$ & $25\pm5$ & $6.5_{-2.0}^{+8}$ & \cite{2001MNRAS.327..435G} \\
\end{longtable}

\subsection{BH masses}

Table \ref{tab:MBH} presents mass estimates for 53 black holes, primarily from X-ray binaries in our Galaxy and several nearby galaxies. In this paper, compared to \cite{WRBH}, we supplement the BH sample with estimates of the black hole mass in the LMC X-1 \cite{LMC_X-1}, LMC X-3 \cite{LMC_X-3}, and BW Cir \cite{Petrov2014} systems. The final sample includes 11 HMXBs, 38 LMXBs, three black holes discovered in wide binaries from observations by the Gaia astrometric satellite \cite{2023MNRAS.518.1057E,2023MNRAS.521.4323E,2024AnA...686L...2G}, and one single black hole discovered from a long gravitational microlensing event \cite{2025ApJ...983..104S}.

\begin{figure}[h]
    \centering
    \includegraphics[width=0.9\linewidth]{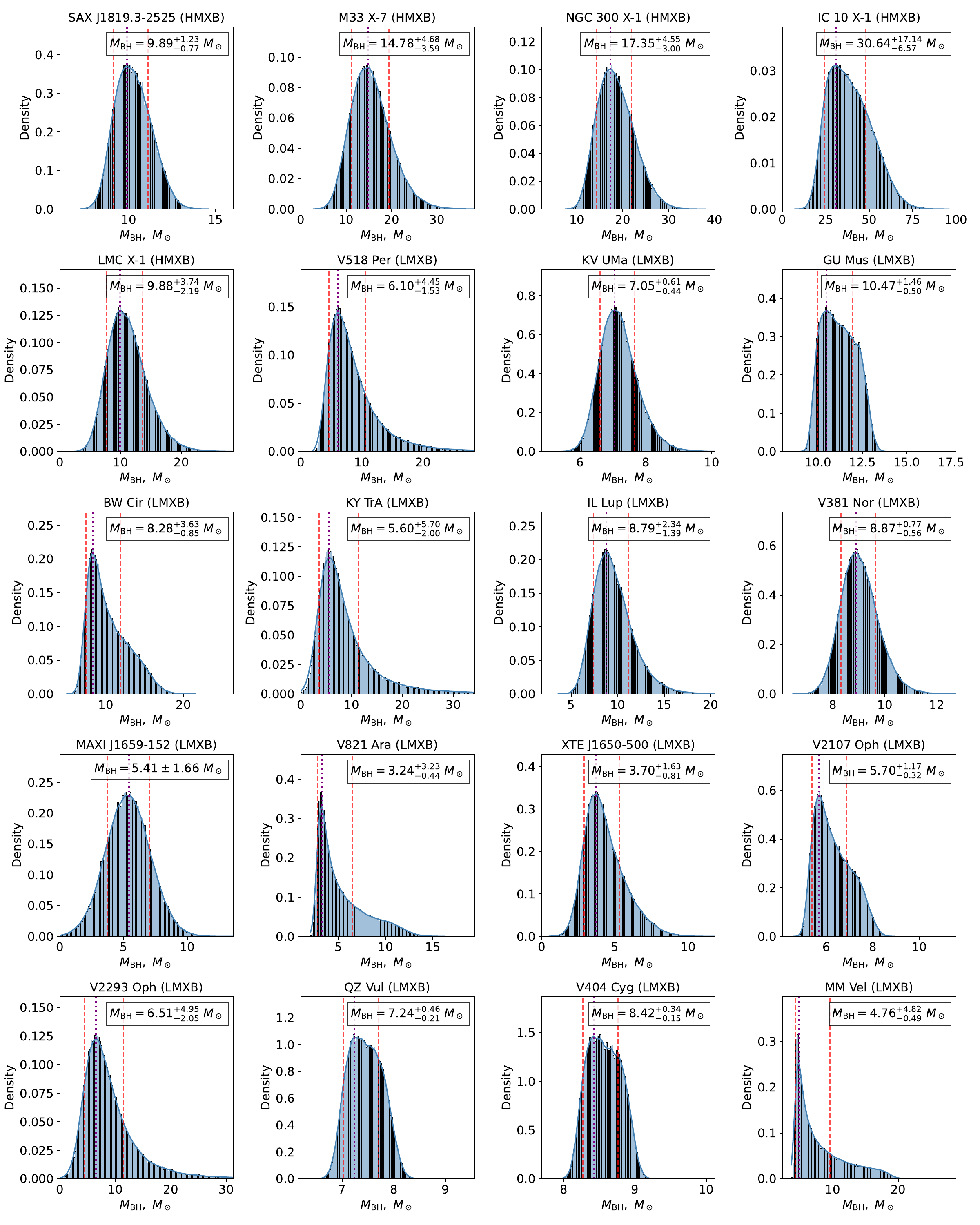}
    \caption{Individual black hole mass distributions for X-ray binaries from Table \ref{tab:MBH} for which the mass function $f_v(M)$, mass ratio $q$, and orbital inclination $i$ are known.}
    \label{fig:BH_individuals}
\end{figure}

The distribution of individual masses in most cases was assumed to be normal with a variance corresponding to the determination error.
If observational estimates of the optical star mass function $f_v(M)$, the mass ratio $q = M_v/M_\mathrm{BH}$, and the orbital inclination $i$ were available in the literature for a given binary system with a black hole, the black hole mass distribution was calculated using the method from \cite{Petrov2014}: based on observational information on the possible distribution of the parameters $f_v(M)$, $q$, and $i$, $10^5$ artificial samples of these parameters were generated, from which the black hole mass was calculated using the formula $M_\mathrm{BH} = f_v(M)(1+q)^2(\sin{i})^{-3}$.
The resulting distributions were smoothed with a filter with a Gaussian kernel and subsequently used to estimate the empirical distribution function. Fig. \ref{fig:BH_individuals} shows individual distributions of $M_\mathrm{BH}$ obtained in this way.

\break 

\begin{figure}
    \centering
    \includegraphics[width=0.9\linewidth]{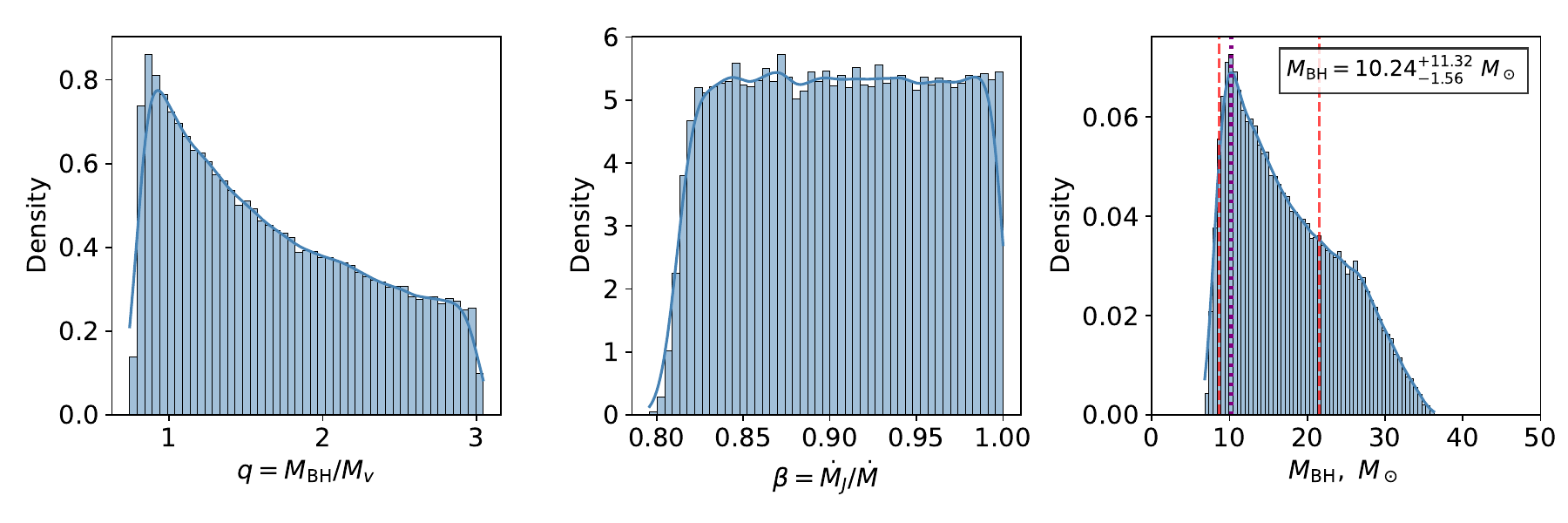}  
    \caption{Distributions of the mass ratio $q$ (left), parameter $\beta$ (center), and the black hole mass (right) in SS 433 from the numerical solution of equation (\ref{eq:SS433_q}).}
    \label{fig:ss433}
\end{figure}

The mass distribution of the black hole in SS 433 was calculated separately. Spectroscopic measurement of the motion of the optical star in this system is difficult, although the orbital inclination angle $i = 78^\circ.88$ is well determined from observations of relativistic jets \cite{Davydov2008}. Based on photometric observations and evolutionary considerations, Goranskij \cite{Goranskij2011} estimated the optical star mass: $M_v = 9-12~M_\odot$. Additional information on the possible mass of the compact object (more precisely, on the mass ratio $q$) in this system can be inferred from the observed increase in the orbital binary period at a rate of $\dot{P} = (1.14\pm0.25)\cdot10^{-7}~\text{s}/\text{s}$ \cite{SS433_Pdot}. From the model of increase in the orbital period due to an isotropic disk wind with possible matter outflow from the system through the circumbinary shell, a relationship follows that connects the mass ratio $q$ with the observed parameters \cite{SS433_ecc}:
\begin{equation}
    \label{eq:SS433_q}
    (3-A)q^2 + (2-A)q - 3\beta - 3K(1-\beta)(1+q)^{5/3} = 0, \quad A = \left( \frac{\dot P_b}{P_b} \right) \cdot \left( -\frac{M_v}{\dot M_v} \right),
\end{equation}
where $P_b$ and $\dot P_b$ are the orbital period of the SS 433 system and its change rate, $M_v$ and $\dot M_v < 0$ are the optical star mass and its mass loss, $0 \leq \beta \leq 1$ is the fraction of $\dot M_v$ that falls within the Jeans mode of mass loss, and $K$ is a dimensionless parameter characterizing the loss of angular momentum through the circumbinary disk. By fixing the values $\dot M_v = -10^{-4}~M_\odot/\text{year}$ \cite{Shkovskii1981}, $P_b = 13.082~\text{day}$ \cite{SS433_Pdot}, $K = 4.7$ \cite{SS433_ecc} and varying $M_v$ (uniformly distributed within $M_v = 9-12~M_\odot$), $\dot P_b = (1.14\pm0.25)\cdot 10^{-7}~\text{s}/\text{s}$ (normally distributed), and $\beta$ (uniformly distributed from 0 to 1), we can estimate the distribution of possible values of $q$ by numerically solving equation (\ref{eq:SS433_q}). Calculations show that equation (\ref{eq:SS433_q}) has a positive real root with respect to $q$ at $\beta = 0.8\div1.0$ and $q \geq 0.8$ in full agreement with the results in paper \cite{SS433_ecc}. The calculated distributions of $q$, $\beta$, and $M_x = M_\mathrm{BH}$ are shown in Fig. \ref{fig:ss433}.

The solution of equation (\ref{eq:SS433_q}) under known observational constraints allows us to unambiguously determine only the lower bound for the mass ratio $q = 0.82$ and the black hole mass $M_\mathrm{BH} = 7.38~M_\odot$ for $\beta=1$ and $M_v = 9~M_\odot$ \cite{SS433_ecc}. On the other hand, for $\beta < 0.8$, equation (1) does not have a positive real solution for $q$. Thus, the main scatter of $q$ and $M_\mathrm{BH}$ in case of SS 433 is due to uncertainty in the coefficient $\beta$, the exact determination of which is impossible within the framework of the used approach.


In addition to SS 433, the secular increase in the orbital period and the corresponding estimate of the mass loss rate from the optical star were used in \cite{2022ApJ...926..123A} to evaluate the mass of a compact object in Cyg X-3.

\begin{longtable}{|c|c|c|c|c|c|c|}
    \caption{Observational estimates of black hole masses.} \label{tab:MBH} 
    \\ \hline System & Type & $f_v(M),~M_\odot$ & $q = M_v/M_\mathrm{{BH}}$ & $i,~^\circ$ & $M_\mathrm{{BH}},~M_\odot$ & Ref. \\ \hline \endfirsthead
    
    \caption{Observational estimates of black hole masses (continued).} 
    \\ \hline System & Type & $f_v(M),~M_\odot$ & $q = M_v/M_\mathrm{{BH}}$ & $i,~^\circ$ & $M_\mathrm{{BH}},~M_\odot$ & Ref. \\ \hline \endhead
    \hline \endfoot
    \hline \endlastfoot
        HD 96670 & HMXB & -- & -- & -- & $6.2\pm0.8$ & \cite{2021ApJ...913...48G} \\
        1E 1740.7-2942 & HMXB & -- & -- & -- & $5.1\pm0.4$ & \cite{2020MNRAS.493.2694S} \\
        SAX J1819.3-2525 & HMXB & $2.74\pm0.12$ & $0.67\pm0.04$ & $60.0\div70.7$ & $9.9_{-0.8}^{+1.2}$ & \cite{2001ApJ...555..489O} \\
        SS 433 & HMXB & -- & -- & -- & $10.2_{-1.6}^{+11}$ & \cite{2023NewA..10302060C} \\
        Cyg X-1 & HMXB & -- & -- & -- & $21.2\pm2.2$ & \cite{2022ApJ...934....4K} \\
        Cyg X-3 & HMXB & -- & -- & -- & $7.2\pm1.0$ & \cite{2022ApJ...926..123A} \\
        M33 X-7 & HMXB & $0.46\pm0.08$ & $4.47\pm0.61$ & $74.6\pm1.0$ & $15_{-4}^{+5}$ & \cite{2022AnA...667A..77R} \\
        NGC 300 X-1 & HMXB & $2.6\pm0.3$ & $1.05\div1.65$ & $60\div75$ & $17_{-3}^{+5}$ & \cite{2021ApJ...910...74B} \\
        IC 10 X-1 & HMXB & $7.64\pm1.26$ & $0.7\div1.7$ & $75\div90$ & $30_{-6}^{+17}$ & \cite{2024ApJ...974..184W} \\
        LMC X-1 & HMXB & $0.148\pm0.004$ & $2.91\pm0.49$ & $36.4\pm2.0$ & $9.9_{-2.2}^{+4}$ & \cite{LMC_X-1} \\
        LMC X-3 & HMXB & -- & -- & -- & $6.98\pm0.56$ & \cite{LMC_X-3} \\
        V518 Per & LMXB & $1.21\pm0.06$ & $0.045\div0.25$ & $36.5\pm5.9$ & $6.1_{-1.5}^{+4}$ & \cite{2024MNRAS.531.4917C} \\
        1A 0620-00 & LMXB & -- & -- & -- & $5.86\pm0.24$ & \cite{2017MNRAS.472.1907V} \\
        MAXI J0637-430 & LMXB & -- & -- & -- & $5.1\pm1.6$ & \cite{2022MNRAS.515.3105S} \\
        KV UMa & LMXB & $6.1\pm0.3$ & $0.012\div0.024$ & $74\pm4$ & $7.1_{-0.4}^{+0.6}$ & \cite{2019MNRAS.490.3287C} \\
        GU Mus & LMXB & $3.02\pm0.06$ & $0.079\pm0.007$ & $43.5\pm2.4$ & $10.5_{-0.5}^{+1.5}$ & \cite{2016ApJ...825...46W} \\
        BW Cir & LMXB & $5.73\pm0.29$ & $0.12\pm0.04$ & $74\pm4$ & $8.3_{-0.9}^{+4}$ & \cite{Petrov2014} \\
        MAXI J1305-704 & LMXB & -- & -- & -- & $8.9\pm1.6$ & \cite{2021MNRAS.506..581M} \\
        MAXI J1348-630 & LMXB & -- & -- & -- & $14.8\pm0.9$ & \cite{2023AnA...669A..57T} \\
        CRTS J135716.8-093238 & LMXB &  & -- & -- & $12.4\pm3.6$ & \cite{2016ApJ...822...99C} \\    
        KY TrA & LMXB & $3.2\pm1.0$ & $0.0\div0.31$ & $57\pm13$ & $5.6_{-2.0}^{+6}$ & \cite{2024MNRAS.527.5949Y} \\
        MAXI J1535-571 & LMXB & -- & -- & -- & $10.4\pm0.6$ & \cite{2019MNRAS.487.4221S} \\
        MAXI J1543-564 & LMXB & -- & -- & -- & $13.0\pm1.0$ & \cite{2016ApJ...827...88C} \\
        IL Lup & LMXB & $0.25\pm0.01$ & $0.25\div0.31$ & $20.7\pm1.5$ & $8.8_{-1.4}^{+2.3}$ & \cite{1998ApJ...499..375O} \\
        V381 Nor & LMXB & $7.73\pm0.4$ & $0.0\div0.04$ & $75\pm4$ & $9.0\pm0.6$ & \cite{2002ApJ...568..845O,Petrov2014} \\
        X Nor X-1 & LMXB & -- & -- & -- & $10.0\pm0.1$ & \cite{2014ApJ...789...57S} \\
        V1033 Sco & LMXB & -- & -- & -- & $5.31\pm0.07$ & \cite{2014MNRAS.437.2554M} \\
        MAXI J1659-152 & LMXB & $4.4\pm1.4$ & $0.02\div0.07$ & $70\div80$ & $5.4\pm1.6$ & \cite{2021MNRAS.501.2174T} \\
        V821 Ara & LMXB & $1.91\pm0.08$ & $0.18\pm0.05$ & $37\div78$ & $3.2_{-0.4}^{+3}$ & \cite{2017ApJ...846..132H} \\
        XTE J1650-500 &	LMXB & $2.73\pm0.56$ & $0.0\div0.1$ & $50\div80$ & $3.7_{-0.8}^{+1.6}$ & \cite{2004ApJ...616..376O,Petrov2014} \\
        V2107 Oph & LMXB & $4.86\pm0.13$ & $0.0\div0.053$ & $60\div80$ & $5.7_{-0.3}^{+1.2}$ & \cite{1997PASP..109..461F,Petrov2014} \\
        $[$KRL2007b$]$ 222 & LMXB & -- & -- & -- & $12.2\pm3.5$ & \cite{2015ApJ...807..108I} \\
        V2293 Oph & LMXB & $4.1\pm1.2$ & $0.0\div0.25$ & $61\pm12$ & $6.5_{-2.1}^{+5}$ & \cite{2023MNRAS.526.5209C} \\
        XTE J17464-321 & LMXB & -- & -- & -- & $11.2\pm1.96$ & \cite{2017ApJ...834...88M} \\
        XTE J1752-223 & LMXB & -- & -- & -- & $9.6\pm0.9$ & \cite{2010ApJ...723.1817S} \\
        2XMM J180112.4-254436 & LMXB & -- & -- & -- & $7.0\pm2.4$ & \cite{2006PhDT........26B} \\
        MAXI J1820+070 & LMXB & -- & -- & -- & $8.48\pm0.79$ & \cite{2020ApJ...893L..37T} \\
        MAXI J1836-194 & LMXB & -- & -- & -- & $8.5\pm3.5$ & \cite{2014MNRAS.439.1381R} \\
        V406 Vul & LMXB &  & -- & -- & $7.8\pm1.9$ & \cite{2022MNRAS.517.1476Y} \\
        MAXI J1910-057 & LMXB & -- & -- & -- & $9.98\pm3.67$ & \cite{2023AdSpR..71.1045N} \\
        Granat 1915+105 & LMXB & -- & -- & -- & $10.1\pm0.6$ & \cite{2013ApJ...768..185S} \\
        V1408 Aql & LMXB & -- & -- & -- & $5.0\pm1.0$ & \cite{2021RAA....21..214S} \\
        QZ Vul & LMXB & $4.97\pm0.1$ & $0.033\div0.038$ & $61\div66$ & $7.24_{-0.21}^{+0.5}$ & \cite{2025ARep...69.1063C} \\
        V404 Cyg & LMXB & $6.08\pm0.06$ & $0.056\div0.063$ & $66\div70$ & $8.42_{-0.15}^{+0.3}$ & \cite{1992ApJ...401L..97W} \\
        MM Vel & LMXB & $3.17\pm0.12$ & $0.0264\pm0.004$ & $37\div80$ & $4.8_{-0.5}^{+5}$ & \cite{2003AnA...404..301R} \\
        MAXI J1728-360 & LMXB & -- & -- & -- & $4.6\pm0.5$ & \cite{2023MNRAS.519..519S} \\
        XTE J1818-245 & LMXB & -- & -- & -- & $4.0\pm0.5$ & \cite{2021ApJ...912..110B} \\
        MAXI J1828-249 & LMXB & -- & -- & -- & $4.0\pm0.5$ & \cite{2019PASJ...71..108O} \\
        EXO 1846-031 & LMXB & -- & -- & -- & $3.24\pm0.20$ & \cite{2020AAS...23515902S} \\
        Gaia BH1 & Gaia BH & -- & -- & -- & $9.62\pm0.18$ & \cite{2023MNRAS.518.1057E} \\
        Gaia BH2 & Gaia BH & -- & -- & -- & $8.94\pm0.34$ & \cite{2023MNRAS.521.4323E} \\
        Gaia BH3 & Gaia BH & -- & -- & -- & $32.7\pm0.82$ & \cite{2024AnA...686L...2G} \\
        OGLE-2011-BLG-0462 & Single & -- & -- & -- & $7.15\pm0.83$ & \cite{2025ApJ...983..104S} \\
\end{longtable}

\begin{figure}
    \centering
    \includegraphics[width=1\linewidth]{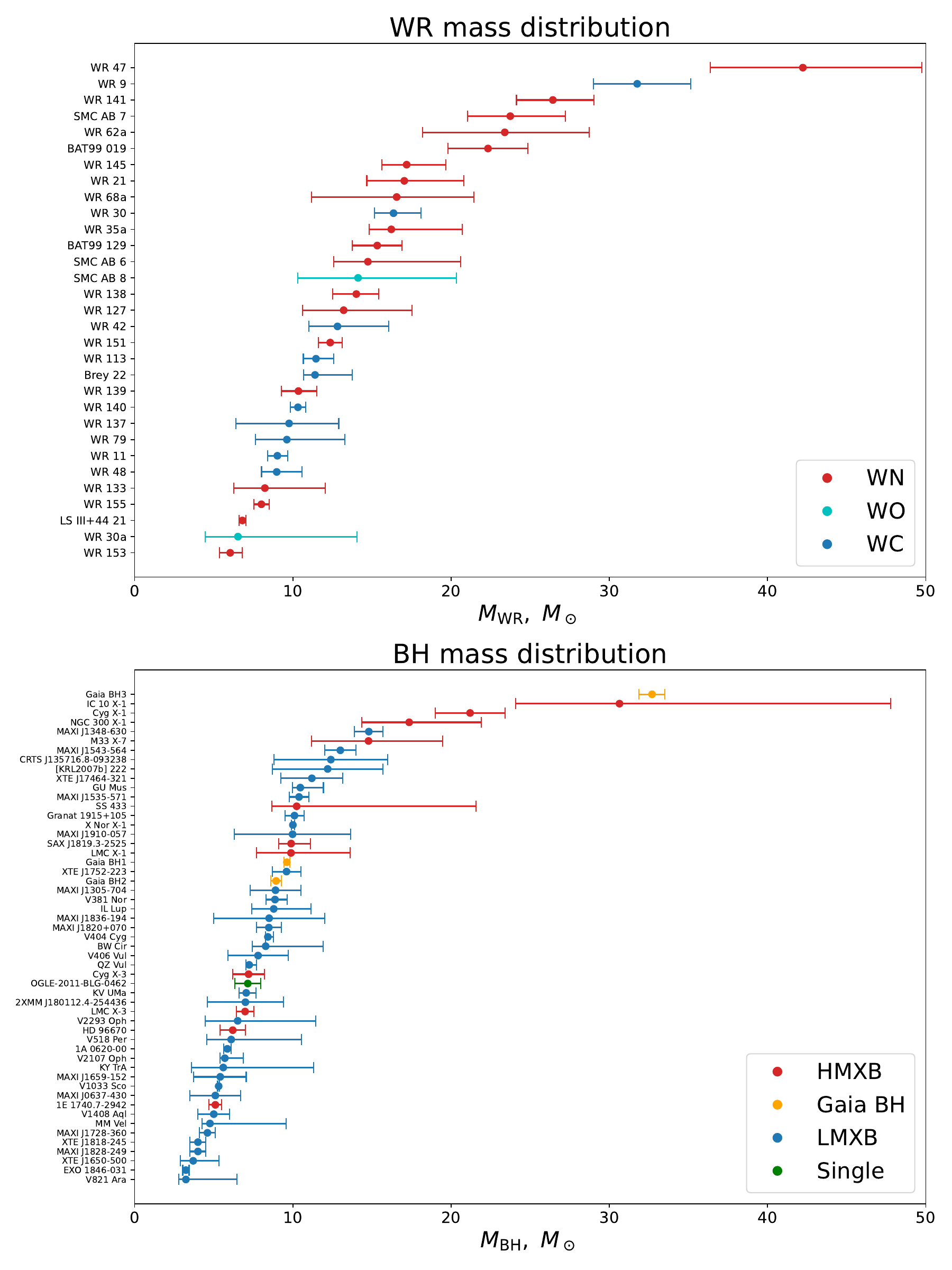}
    \caption{Observed masses of WR stars in WR+OB binaries (top panel) and black holes in binaries (bottom panel). Objects of different subtypes are shown in different colors.}
    \label{fig:WRBH_data}
\end{figure}

Fig. \ref{fig:WRBH_data} presents available observational data on the masses of WR stars in WR+OB binaries and the masses of black holes in binary systems (primarily in X-ray binaries). Different colors in the Figure indicate objects of different types — WRs are divided by their spectral subclasses, and BHs by the type of binary system.

\section{Accounting for mass loss at the WR stage}
\label{sec:dot_MWR}

WR stars exhibit a significant stellar wind mass loss ($\dot M_\mathrm{WR} \sim 10^{-5}~M_\odot/\text{yr}$), which can lead to a difference of up to several solar masses between the masses of a WR star at the beginning of this stage and at its end, i.e., immediately prior to the WR core collapse into a black hole. Thus, to correctly compare the WR and BH masses it is necessary to take into account the difference between the observed WR masses and their expected masses at the end of the WR stage.

We assume that the total evolutionary time of a star at the WR stage is determined by its final mass. It can be estimated using the formula (see \cite{2022ARep...66S.567C})\begin{equation}
    \label{eq:TWR}
    T = T(M_{\mathrm{WR},f}) = C_T / \sqrt{M_{\mathrm{WR},f}},\quad C_T \approx 1.74\cdot10^6~[\text{yr}]\cdot M_\odot^{1/2}.
\end{equation}
With a power-law dependence of the WR mass loss rate  on its current mass \cite{1989A&A...220..135L}
\begin{equation}
    \label{eq:Mdot}
    \dot M_{\mathrm{WR}} = k M_{\mathrm{WR}}^\alpha, \quad k<0
\end{equation}
the dependence of the final mass of the WR star $M_{\mathrm{WR},f}$ on the initial mass at this stage $M_{\mathrm{WR},i}$ has the form
\begin{equation}
    \label{eq:MWR_final}
    {M_{\mathrm{WR},f}} = ({M_{\mathrm{WR},i}}^{1-\alpha} + k(1-\alpha)T)^{\frac{1}{1-\alpha}}.
\end{equation}

Parameters of the dependence (\ref{eq:Mdot}) can be deduced by comparing observational estimates of the WR mass loss rate of WR stars $\dot M_{\mathrm{WR}}$ with their masses. $\dot M_{\mathrm{WR}}$ can be most reliably determined from measurements of the secular variation of the orbital periods $\dot P$ of WR+OB close binary systems. To date, $\dot P$ has been measured for five systems of this type, for four of which (WR 151, WR 139, WR 127, and WR 141) observational data allow us to estimate $\dot M_{\mathrm{WR}}$. The formula relating $\dot P$ and $\dot M_{\mathrm{WR}}$ with account of the finite size of a WR star reads \cite{2023MNRAS.523.1524S}
\begin{equation}
    \label{eq:dP-dM}
    \frac{\dot P}{P}=-\frac{2\dot M_\mathrm{WR}}{M_\mathrm{WR}}\left\{
    \frac{q}{1+q} - (1+q)\left( \frac{R_\mathrm{WR}}{a} \right)^2 
    \right\}.
\end{equation}
Here $q = M_\mathrm{WR}/M_\mathrm{OB}$, $R_\mathrm{WR}$ is the radius of the WR star (it can be estimated based on the solution of light curves of eclipsing systems or from model considerations), $a = a_\mathrm{WR}+a_\mathrm{OB}$ is the orbital binary separation. The WR stellar wind is assumed to be isotropic, the orbit of the system is circular, and the influence of the OB-star stellar wind on the orbital evolution is neglected. The correction for the radius turns out to be significant only in the case of WR 151 (about 40\%). For the WR 139 system, the corresponding correction does not exceed 4\%, and in the case of sufficiently wide systems WR~127 and WR 141, the influence of the radius of the WR star on the orbital evolution can be neglected.

\begin{figure}[h]
    \centering
    \includegraphics[width=0.9\linewidth]{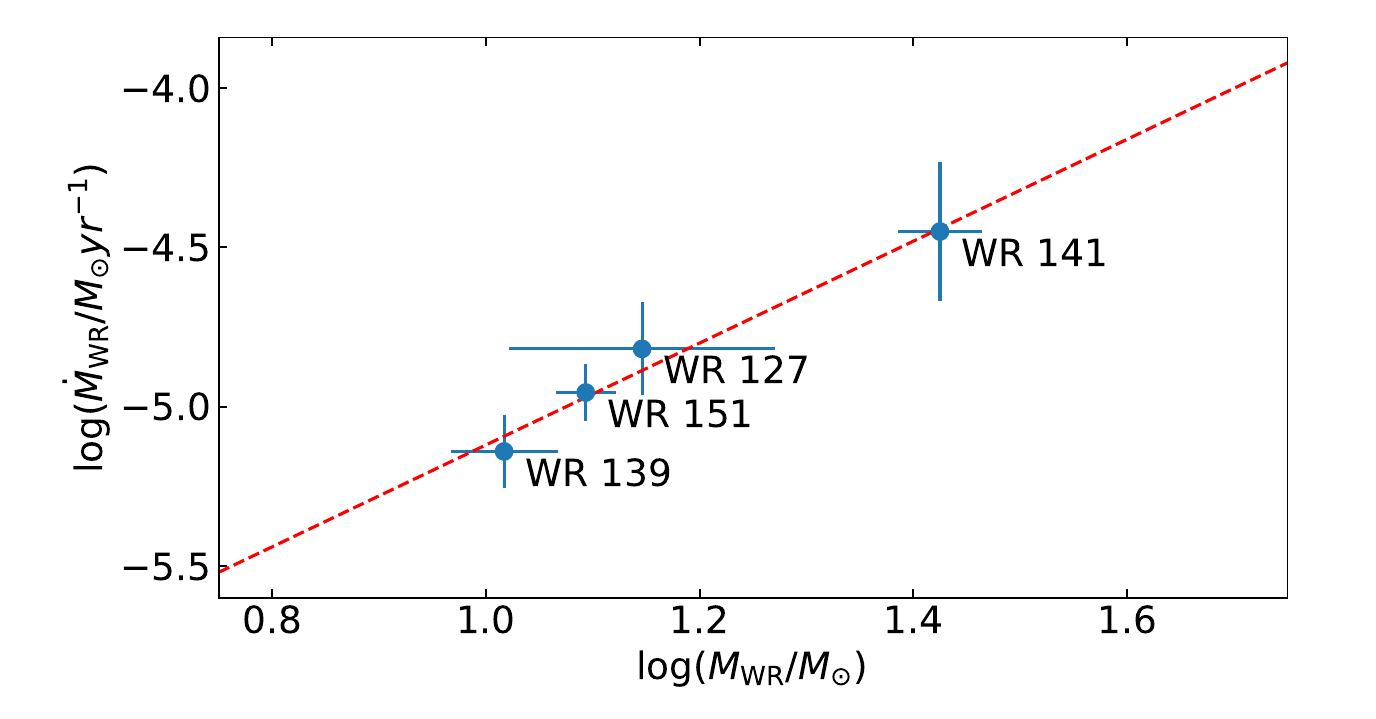}
    \caption{Empirical dependence of the mass loss rate of the WR star $\dot M_{\mathrm{WR}}$ on its mass $M_{\mathrm{WR}}$.}
    \label{fig:dMWR-MWR}
\end{figure}

Table \ref{tab:dMWR-MWR} lists the WR masses in the WR 151, WR 139, WR 127, and WR 141 systems and the mass loss rates $\dot M_{\mathrm{WR}}$ calculated from formula (\ref{eq:dP-dM}). Fig. \ref{fig:dMWR-MWR} plots the dependence $\dot M_{\mathrm{WR}}(M_{\mathrm{WR}})$ in logarithmic coordinates and the best-fit power-law dependence (\ref{eq:Mdot}). The resulting empirical relation has the form:
\begin{equation}
    \label{eq:dMWR-MWR_obs}
    \log_{10}(\dot M_\mathrm{WR}/M_\odot~\text{yr}^{-1}) = (1.60\pm0.13)\cdot\log_{10}(M_\mathrm{WR}/M_\odot) - (6.72\pm0.15),
\end{equation}

\break

\begin{table}[h!]
    \centering
    \caption{Estimates of the WR star mass and WR star mass loss rate in the WR 151, WR 139, WR 127, and WR 141 systems.}
    \label{tab:dMWR-MWR}
    \begin{tabular}{|c|c|c|c|c|c|c|}
        \hline
        System & $M_\mathrm{WR},~M_\odot$ & $P$, days & $\dot P$, s/yr & $R_\mathrm{WR},~R_\odot$ & $\dot M_\mathrm{WR},~M_\odot~\text{yr}^{-1}$ & Ref. \\
        \hline
        WR 151   & $12.4\pm0.8$ & 2.126937$\pm$0.000004 & 0.077$\pm$0.010 & $\sim7.3$ & $(1.11\pm0.22)\cdot10^{-5}$ & \cite{2023MNRAS.523.1524S,2009PASP..121..708H} \\
        WR 139 & $10.4\pm1.2$ & 4.2125319$\pm$0.0000015 & 0.134$\pm$0.03 & $\sim3$ & $(7.2\pm1.9)\cdot10^{-6}$ & \cite{2023MNRAS.526.4529S,2011AN....332..616E,Antokhin2026_V444} \\
        WR 127 & $14\pm4$   & 9.55492$\pm$0.00005 & 0.62$\pm$0.11 & -- & $(1.5\pm0.5)\cdot10^{-5}$       & \cite{2024AnA...683L..17S} \\
        WR 141 & $26.6\pm2.4$ & 21.6897$\pm$0.0005 & 2.5$\pm$0.8 & -- & $(3.5\pm1.7)\cdot10^{-5}$     & \cite{2024ARep...68.1145S} \\
        \hline
    \end{tabular}
\end{table}

Even with the known dependence (\ref{eq:dMWR-MWR_obs}), a large uncertainty is introduced by the time $T_\mathrm{WR}$ a given star spends in the WR stage before the core collapse, or more precisely, the ''age'' of the WR star at this stage. Unfortunately, determining the age of each star is problematic, but certain assumptions can be made about the average age of observed WR stars. In two extreme cases the observed stars are at the very beginning of the WR stage, i.e., the mass loss for all stars before collapse will occur for the maximum time (\ref{eq:TWR}), or all observed actual WR masses correspond to their final masses. An intermediate assumption is that all observed WR stars are in the middle of the stage, i.e., the mass loss before the core collapse in each case occur during half the time from formula (\ref{eq:TWR}). These three possibilities and the resulting relationships between the final WR masses and the BH masses will be discussed in the following sections.

\break

\section{The WR and BH mass distribution functions}

In previous studies, such as \cite{WRBH} and \cite{Petrov2014}, mass distribution functions were calculated using a simple sum of the individual distribution functions of the sample objects. In the present paper, the distribution function is estimated directly using the individual mass distributions. For a given mass $M$, $p_i(M_{obj} < M) \equiv p_i$ is calculated for each sample element, i.e., the value of the cumulative distribution function of the individual distribution at point $M$. Then, for each $N = 0,1,...N_{type}$ ($N_{type}$ is the number of the full sample of objects of a given type), the probability $P(N)$ that out of $N_{type}$ objects, $N$ will have a mass less than $M$ is calculated. This statistic corresponds to the probability distribution that the value of $M$ corresponds to the $p = (N/N_{type})$-th quantile of the overall distribution, and follows the Poissonian binomial distribution
\begin{equation}
    \label{eq:PBD}
    P(N) = \sum_{A \in F_N} \prod_{i \in A} p_i \prod_{j \in A^c} (1-p_j),
\end{equation}
where $F_N$ is the set of all ordered subsets of $N$ numbers selected from the series 1,2,...$N_{type}$, $A^c$ is the complement of the set $A$. For relatively small numbers $N$, this statistic is calculated quickly.

By performing this procedure for a set of $M$ values, we estimate the mass distribution function $P(M, p)$ for objects of this type.
To clearly identify regions on the plots that are most densely covered by observational data, it is appropriate to normalize the obtained WR and BH mass distribution estimates by the distribution quantile $p$-values along the $M$ mass axis.
For further calculations, the estimated WR and BH mass distributions are additionally interpolated onto a common grid of $p$-values.

A convenient method for comparing distributions is the so-called quantile-quantile plot (Q-Q plot). If the transformation $M_\mathrm{WR} \rightarrow M_\mathrm{OB}$ is monotonic, the quantile-quantile plot will follow the function of this transformation.
The WR and BH distributions constructed in the way described above allow us to produce the quantile-quantile plot taking into account the uncertainty that smears it.

\begin{figure}
    \centering
    \includegraphics[width=1\linewidth]{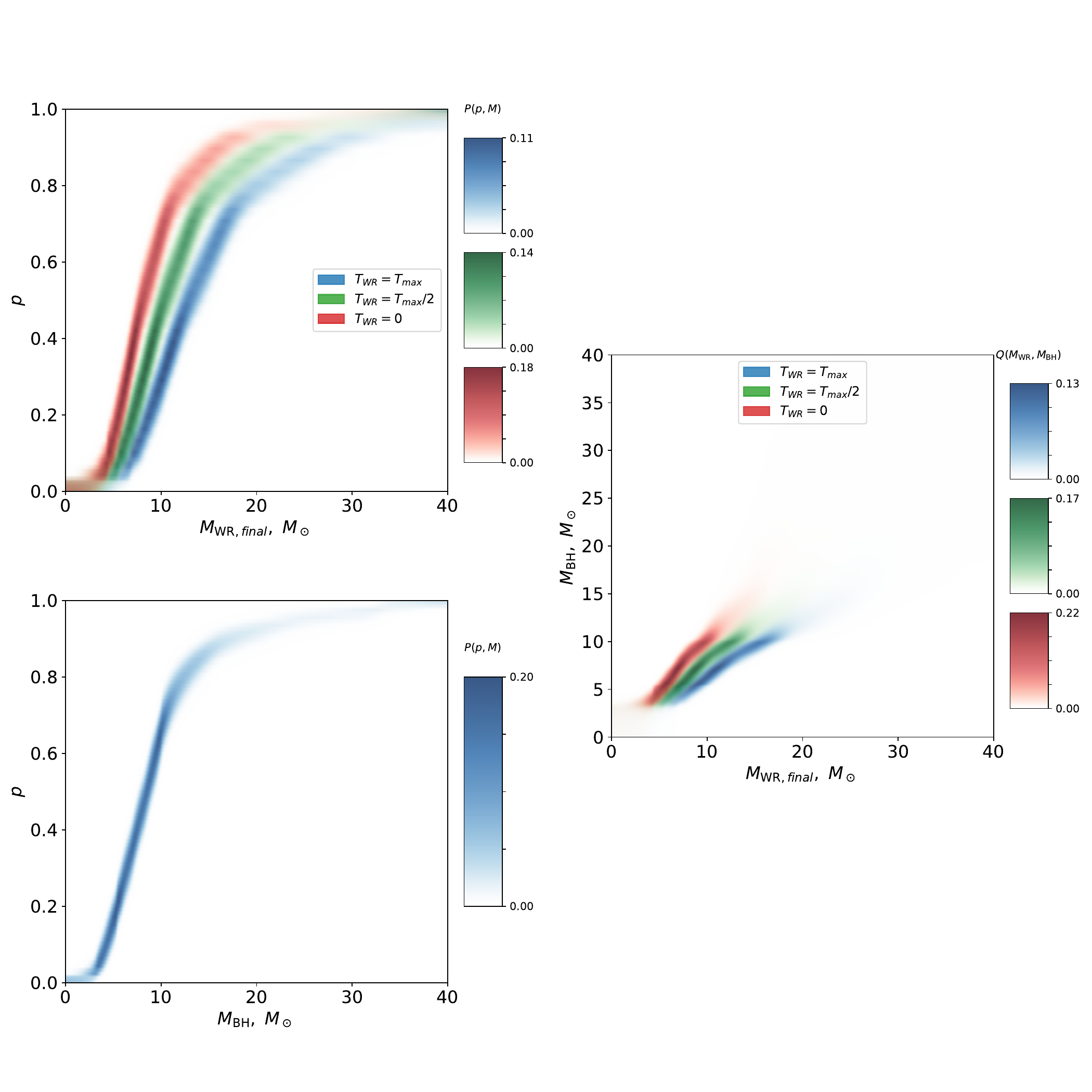}
    \caption{Observational mass distribution functions for WR (upper left panel) and BH (lower left panel) stars. Three different colors in the WR and $Q(M_\mathrm{WR},M_\mathrm{BH})$ distribution plots show different mass loss treatment during the WR stage. On the right are quantile-quantile plots of the WR and BH mass distributions for three WR mass loss  models. The color encodes the set of probabilities normalized along the vertical axis estimated by formula (\ref{eq:PBD}) (in the plots on the left), and the result of the matrix product of the normalized estimates of the $M_\mathrm{WR}$ and $M_\mathrm{BH}$ distributions by formula (\ref{eq:Q}) (on the right).}
    \label{fig:WR-BH-QQ}
\end{figure}

Fig. \ref{fig:WR-BH-QQ} shows the calculated mass distributions of WR stars (upper left panel), black holes (bottom right panel), and quantile-quantile plots (right panel).
The WR mass distribution and its comparison with the black hole mass distribution are performed for three mass loss scenarios: the WR mass is constant (mass loss time $T_\mathrm{WR} = 0$), the WR star's mass is lost over the full duration of the stage ($T_\mathrm{WR} = T_{max}$ where $T_{max}$ is determined from expression (\ref{eq:TWR})), and the WR mass is lost over half the full duration of the stage ($T_\mathrm{WR} = T_{max}/2$).
The initial calculation of the distribution functions was performed on an array of mass values from 0 to $50~M_\odot$ with a step of $0.16~M_\odot$ (251 points). The size of the probability array was specified by the number of sample elements $N_{type}$. Afterwards, both distribution plots for WR and BH were interpolated to a common array of probabilities $p$ from 0 to 1 with a step of 0.01 (101 points).
The color in the plots encodes the value of the depicted function in pixels on the used grid along the vertical and horizontal axes (color scales in Fig. \ref{fig:WR-BH-QQ}).
The quantile-quantile plots of the final mass distributions of WR stars and black holes $Q(M_\mathrm{WR},M_\mathrm{BH})$ were constructed by matrix multiplication of the obtained distribution plots along the common probability axis:
\begin{equation}
    \label{eq:Q}
    Q(M_{\mathrm{WR},f},M_\mathrm{BH}) = P(M_\mathrm{BH},p)^T \cdot P(M_{\mathrm{WR},f},p).
\end{equation}

\clearpage
\section{Discussion}

\begin{figure}
    \centering
    \includegraphics[width=0.8\linewidth]{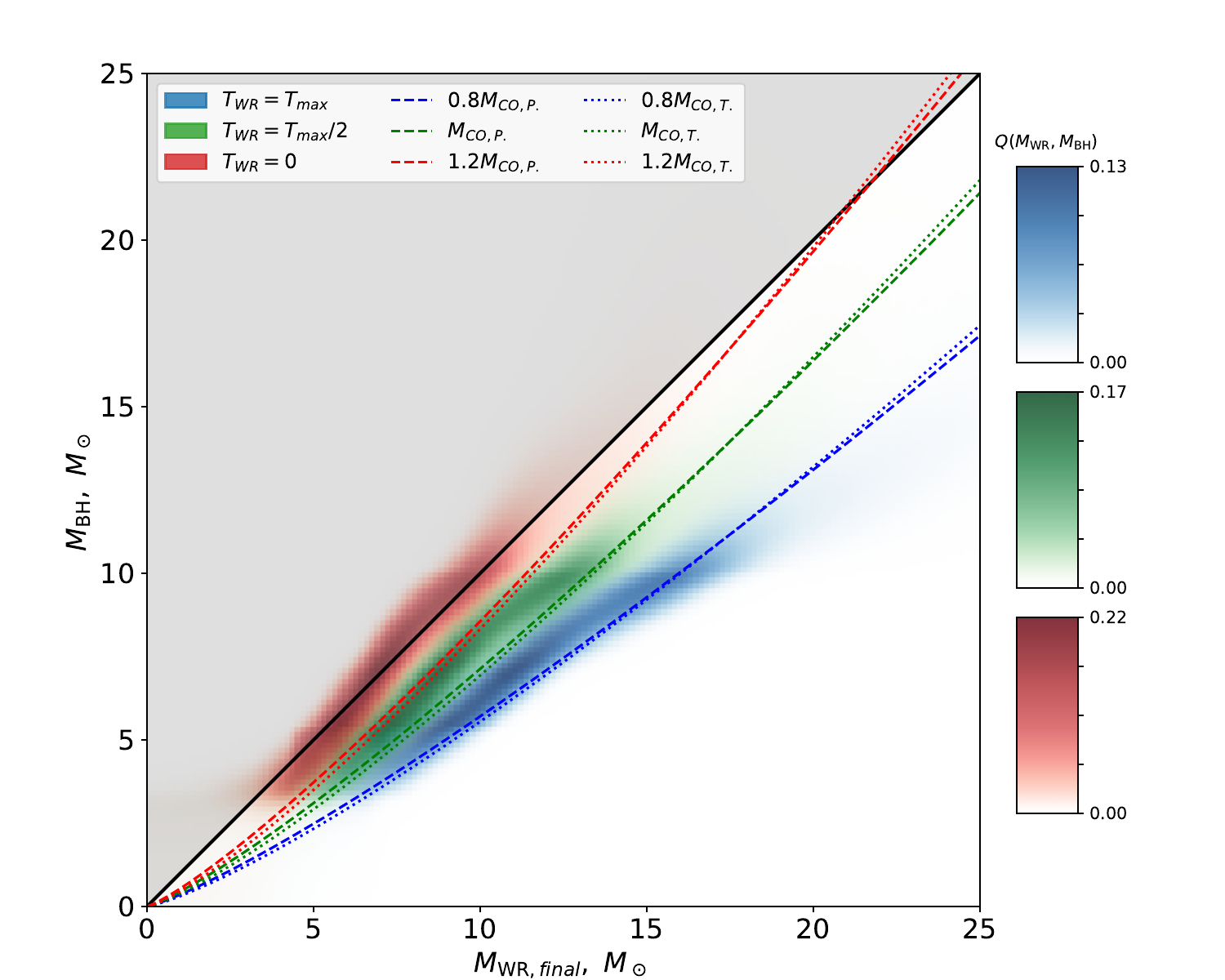}
    \caption{Quantile-quantile plots of the WR and BH mass distributions for three variants of the WR mass loss within 25 $M_\odot$. Dashed lines show the dependences of the mass of the CO core of the WR star on its mass according to ($P. = \text{Paczynski, 1971}$ \cite{1971AcA....21....1P} and  $T. = \text{Takahashi et al., 2023}$ \cite{2023ApJ...945...19T}) with coefficients of 0.8, 1, and 1.2. The black solid line corresponds to the case $M_\mathrm{WR} = M_\mathrm{BH}$, the ''forbidden'' region $M_\mathrm{WR} < M_\mathrm{BH}$ is shaded in gray.}
    \label{fig:QQ-MCO}
\end{figure}

Fig. \ref{fig:QQ-MCO} shows the same quantile-quantile plots of the WR and BH mass distributions for three mass-loss assumptions for WR stars as in Fig. \ref{fig:WR-BH-QQ}, but for a smaller range of WR and BH masses corresponding to observational mass estimates.
This plot also shows the theoretical dependences of the CO-core mass of a helium (WR) star 
on its total mass with coefficients of 0.8, 1, and 1.2. For this purpose, two well-known theoretical power-law approximations were used: $M_\mathrm{CO} = 0.45M_\mathrm{WR}^{1.20}$ from the classical paper by Paczynski \cite{1971AcA....21....1P} and the recent version $M_\mathrm{CO} = 0.39M_\mathrm{WR}^{1.25}$ obtained from calculations by Takahashi et al. \cite{2023ApJ...945...19T}. The range $\pm 20 \%$ to the mass of the CO nucleus was taken based on the results of our previous work \cite{WRBH} (see formulas 5 and 8 from that paper).
The difference between the current version and the previous one, namely, a stronger mass loss at the WR stage (i.e., a shift of the quantile-quantile graphs to the left relative to the theoretical curves) is associated with a somewhat sharper dependence $\dot M_{\mathrm{WR}}(M_{\mathrm{WR}})$: in paper \cite{WRBH}, the dependence (\ref{eq:Mdot}) with the coefficient $\alpha = 1.5$ was considered, for which the solution of the equation (\ref{eq:MWR_final}) shows a simple proportionality between the initial and final WR masses. With refined coefficients (\ref{eq:dMWR-MWR_obs}), under the assumption of maximum mass loss, most of the quantile-quantile graph ends up in the unphysical region $M_\mathrm{WR} \leq M_\mathrm{BH}$. The quantile-quantile plot for the case without mass loss is located near the curve $M_\mathrm{BH} \approx 0.8 M_\mathrm{CO}$, which is consistent with the results of \cite{WRBH}. The position of the quantile-quantile plot for the average ''age'' of all observed WR stars corresponds to the range $M_\mathrm{BH} \approx (1.0\div1.2) M_\mathrm{CO}$.

\section{Conclusion}
In this paper we analyzed the relationship between the masses of Wolf-Rayet stars and black holes based on available data on WR star masses in WR+OB spectroscopic binaries (Table 1) and black hole masses in binaries in our own and nearby galaxies, or estimated from gravitational lensing in the Galaxy (Table 1). The mass distributions are based on a direct method of estimating the distribution function from a sample of mass values with individual distributions, without using analytical assumptions, as in the \cite{WRBH}. A more refined analysis using a larger sample of objects and more accurate statistical methods than those employed by \cite{WRBH} confirms the evolutionary relationship between the masses of black holes and the CO-cores of Wolf-Rayet stars at stages preceding the collapse, namely, $M_\mathrm{WR}\simeq M_\mathrm{CO}$ (Fig. \ref{fig:QQ-MCO}).
The found empirical relationship is consistent with theoretical expectations \cite{1971AcA....21....1P,2023ApJ...945...19T} without relying on model assumptions about the collapse physics and can be confidently used in massive population synthesis calculations for modeling binary systems with black holes.

\section*{FUNDING}
The study was conducted under the state assignment of Lomonosov Moscow State University.

\section*{CONFLICT OF INTEREST}
The authors of this work declare that they have no conflicts of interest.


\end{document}